\documentclass[aps,prx,superscriptaddress,nofootinbib,twocolumn]{revtex4-1}

\usepackage{mathrsfs}
\usepackage{amsfonts}
\usepackage{amssymb}
\usepackage{amsmath}
\usepackage{graphicx}
\usepackage[usenames,dvipsnames]{color}
\usepackage[colorlinks=true,citecolor=blue,linkcolor=magenta]{hyperref}
\usepackage{ulem}
\usepackage{lmodern}

\newcommand{\figpath}{.}

\newcommand{\ket}[1]{\vert{ #1 }\rangle}

\newcommand{\SIG}[2]{\sigma^{#1}_{#2}}

\begin{document}

\title{One-dimensional quantum computing with a `segmented chain'  \\ is feasible with today's gate fidelities }

\author{Ying Li}

\affiliation{Department of Materials, University of Oxford, Parks Road, Oxford OX1 3PH, United Kingdom}

\author{Simon C. Benjamin}

\affiliation{Department of Materials, University of Oxford, Parks Road, Oxford OX1 3PH, United Kingdom}

\date{\today}

\begin{abstract}
In principle a 1D array of nearest-neighbour linked qubits is compatible with fault tolerant quantum computing. However such a restricted topology necessitates a large overhead for shuffling qubits and consequently the fault tolerance threshold is far lower than in 2D architectures. Here we identify a middle ground: a 1D segmented chain which is a linear array of segments, each of which is a well-connected zone with all-to-all connectivity. The architecture is relevant to both ion trap and solid-state systems. We  establish that fault tolerance can be achieved either by a surface code alone, or via an additional concatenated four-qubit gauge code. We find that the fault tolerance threshold is $0.12\%$ for 15-qubit segments, while larger segments are superior. For 35 or more qubits per segment one can achieve computation on a meaningful scale with today's state-of-the-art fidelities without the use of the upper concatenation layer, thus minimising the overall device size.
\end{abstract}

\maketitle  

\section{Introduction}
\label{sec:Introduction}

Quantum computation can solve some problems that are intractable for classical computation, e.g.~the quantum Shor's algorithm can solve the integer factorization problem in polynomial time while the best known classical algorithm runs in exponential time~\cite{Nielsen2010}. Implementing such quantum algorithms, we need a quantum computer that contains a plenty of qubits~\cite{Fowler2012, Joe2016} with the noise suppressed to the sub-threshold regime~\cite{Szkopek2006, Stephens2009, Fowler2009, Wang2011}. These qubits must be coupled by controllable interactions to form a network. The connectivity of the network is higher, the quantum computer can tolerate more errors, i.e.~the noise threshold is higher. For example, when qubits form a one-dimensional (1D) array with nearest-neighbouring (NN) interactions, the error-rate threshold is between $10^{-7}$---$10^{-5}$ per gate~\cite{Szkopek2006, Stephens2009}; when qubits form a two-dimensional (2D) array with NN interactions, the threshold is about $1\%$ per gate~\cite{Fowler2009, Wang2011}. In this paper, we study the fault-tolerant quantum computing in a 1D array of qubit with short-range but more than NN interactions that exist in many quantum systems, and we show that the threshold could be higher than $0.1\%$ per gate with feasible interaction ranges.

\begin{figure*}[tbp]
\centering
\includegraphics[width=1\linewidth]{\figpath 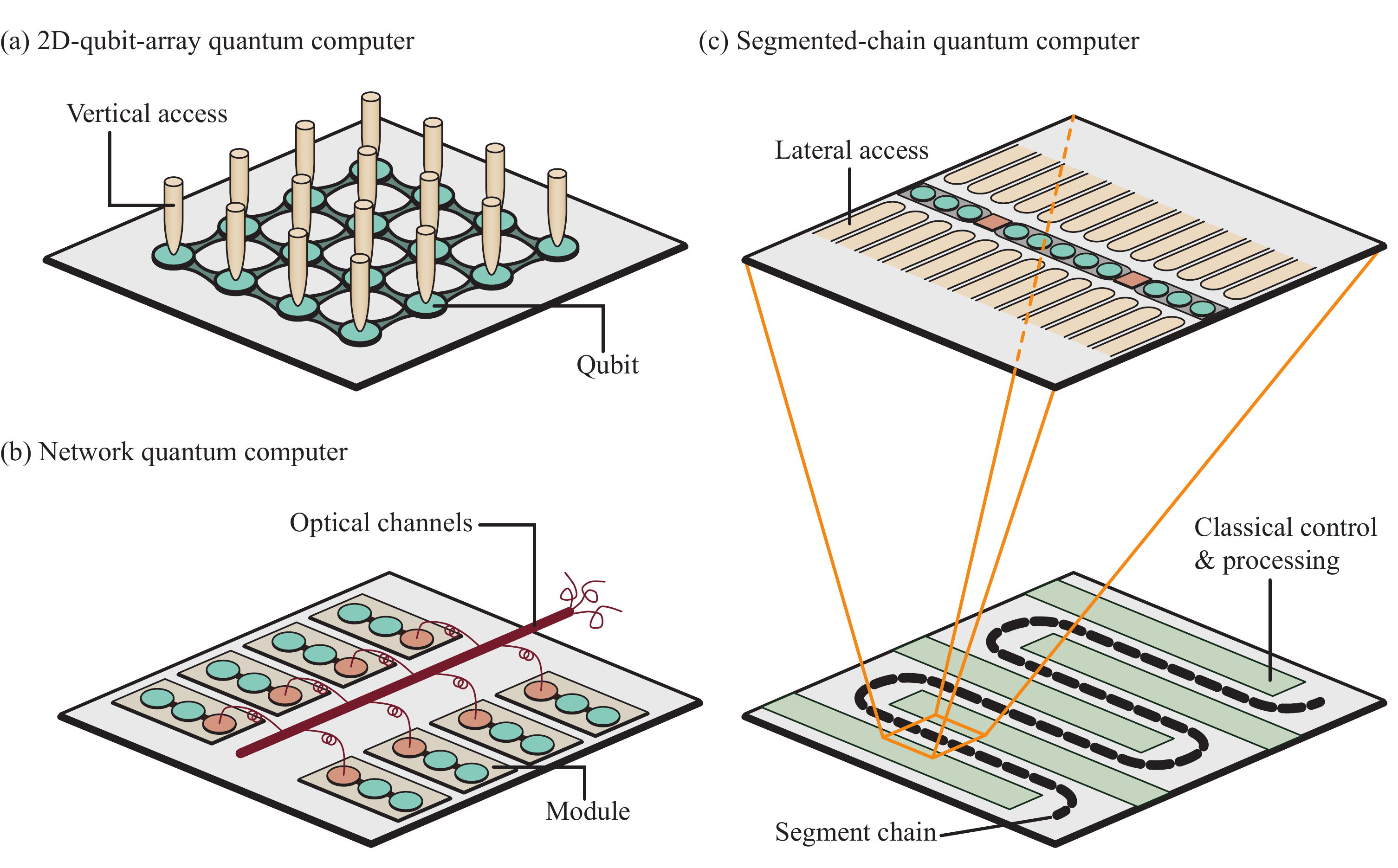}
\caption{
Quantum computers of (a) a 2D qubit array with vertical accesses, (b) a network of modules linked by optical channels and (c) a 1D qubit array with the segmented chain structure. Here we depict a snaking path, to illustrate that a 1D array can efficiently use a 2D surface.
}
\label{fig:architectures}
\end{figure*}

The advantage of using a 1D array of qubits as a quantum computer is that the system could be monolithic and embedded in a 2D surface. Gate error rates lower than $1\%$ have been demonstrated in 1D qubit arrays using ion traps~\cite{Lucas, Wineland} and superconducting qubits~\cite{Barends2014}. These error rates are sub-threshold for the fault-tolerant quantum computing in a 2D qubit array~\cite{Wang2011}, therefore one way of building quantum computers is extending the qubit array to 2D and preserving low error rates at the same time. For ion traps, an approach of scaling up the quantum computer is using the network paradigm, i.e.~ion traps are networked by using quantum communications via optical systems [Fig.~\ref{fig:architectures}(a)]~\cite{Li2012, Nickerson2013, Monroe2014}. The ion-trap network can also be integrated on a chip, in which communications are achieved by using ion transport through junction structures~\cite{Lekitsch2017}. For superconducting qubits, a 2D qubit array can be fabricated on a 2D surface, and then one can access qubits vertically from the third dimension [Fig.~\ref{fig:architectures}(b)]~\cite{Bejanin2016}. As we will show in this paper, a 1D qubit array can also tolerate a high level of noise. Therefore we can also build a quantum computer without optical communication channels, and qubits are accessed laterally using control and read-out lines in the same surface [Fig.~\ref{fig:architectures}(c)]. Although 2D qubit arrays have some advantages, e.g.~a lower cost of communications within the qubit array, 1D qubit arrays may be more adaptable to harsh environments and easier to optimise to suppress noise because of their structural simplicity.

\begin{figure}[tbp]
\centering
\includegraphics[width=1\linewidth]{\figpath 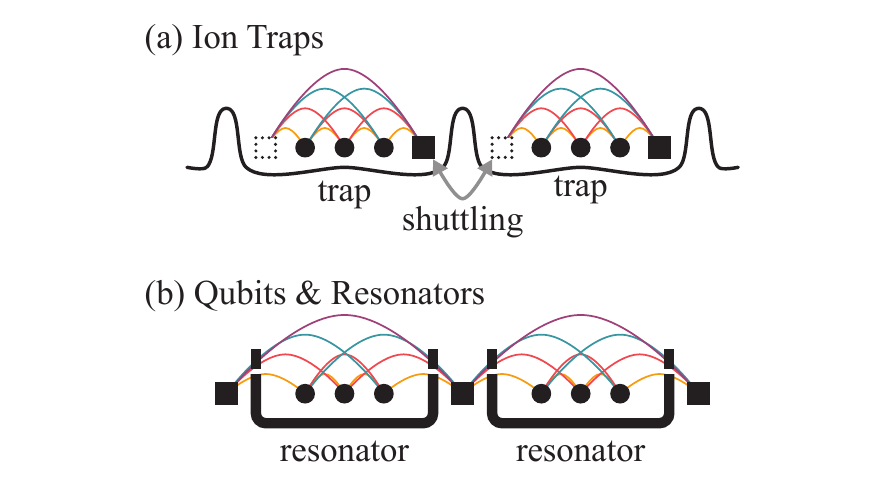}
\caption{
(a) A chain of ion traps. In each trap, qubits (ions, represented by circles and squares) are coupled to the same phonon modes, and two-qubit entangling gates (e.g.~CNOT gates, represented by colour curves) can be performed on any pair of qubits. Traps are coupled by moving qubits (squares) between nearest neighbouring traps. (b) A chain of resonators. Similar to ion traps, in each resonator, qubits (e.g.~superconducting qubits) are coupled to the same photon mode, and two-qubit entangling gates can be performed on any pair of qubits. Two nearest neighbouring resonators are coupled by sharing a qubit.
}
\label{fig:system}
\end{figure}

Medium-range interactions in a 1D array, as required by our architecture, do exist in many quantum systems. In ion traps, ions in the same trap are coupled to common phonon modes. Mediated by such modes, an entangling two-qubit gate can be directly performed on any pair of qubits, i.e.~qubits in the same trap are all-to-all connected [Fig.~\ref{fig:system}(a)]~\cite{Choi2014, Debnath2016}. Using such ion traps as building blocks, the quantum computer is formed by a series of ion traps, in which two NN traps can be coupled by shuttling an ion between two traps~\cite{Kielpinski2002, Rowe2002}. By a different physical mechanism, our requirement may also be achieve with superconducting qubits: when coupled to the same resonator they can also be all-to-all connected~\cite{Paik2016, Takita2016}, and two NN resonators can be coupled using a qubit interacting with both resonators~\cite{Takita2016} [Fig.~\ref{fig:system}(a)]. In both platforms, the quantum computer is a chain of sub-systems, i.e.~segments. Each segment is an ion trap or a resonator. Qubits within a segment are all-to-all connected; and two NN segments are coupled by a shared qubit, which is connected to all qubits in both segments. We find that, if there are enough qubits in each segment, a high level of noise is tolerable. The number of qubits in each segment, i.e.~the interaction range, is fixed and does not scale with the overall size of the quantum computer.

Our protocol for quantum error correction is based on the surface code~\cite{Dennis2002}, with an optional additional level of encoding if the surface code alone proves insufficient. Using the interaction structure in the segmented chain qubit array, the surface code can be efficiently implemented, but its code distance is limited by the number of qubits in each segment.

Note that because qubits in the same segment may need to be operated sequentially, a surface-code error detection cycle in the segmented chain qubit array may be slower (in terms of the circuit depth) than a 2D qubit array by a factor determined by the segment size.

Logical qubits encoded in the surface code form a 1D array with NN interactions, i.e.~each surface-code qubit can only directly talk to two NN surface-code qubits. Error rates for logical gates on surface-code qubits are of course determined by error rates of physical qubits and the limited code distance. If error rates of surface-code qubits are low enough, a quantum algorithm can be directly implemented using surface-code qubits; otherwise, we need to combine the surface code with another code above it to further correct errors. We choose the concatenated 1D four-qubit gauge code as the higher-level code~\cite{Stephens2009}.

The noise threshold of our protocol depends on the segment size. Our numerical results suggest that given a physical error rate $\sim 0.1\%$ per gate, which has been demonstrated in ion traps~\cite{Lucas, Wineland}, and $35$ qubits in each segment, on average $10^{15}$ CNOT gates can be performed on surface-code qubits before a logical error occurs, in which case the concatenation with the gauge code is not required for implementing many quantum algorithms. When the additional concatenation is indeed used then one can use segments of any size greater than 4 to suppress logical errors arbitrarily, provided that the physical error rate is below a certain threshold. We determine this threshold curve, finding for example the threshold error rate is $0.12\%$ when the segment size is 15.

The paper is organised as follows: In Sec.~\ref{sec:surface}, we focus on the surface code. In Sec.~\ref{sec:gauge}, we focus on the concatenation with the gauge code. In Sec.~\ref{sec:FTQC}, we discuss the performance of the whole error-correction protocol. In Sec.~\ref{sec:Conclusions}, we conclude results of this paper.

\section{Surface code}
\label{sec:surface}

\begin{figure}[tbp]
\centering
\includegraphics[width=1\linewidth]{\figpath 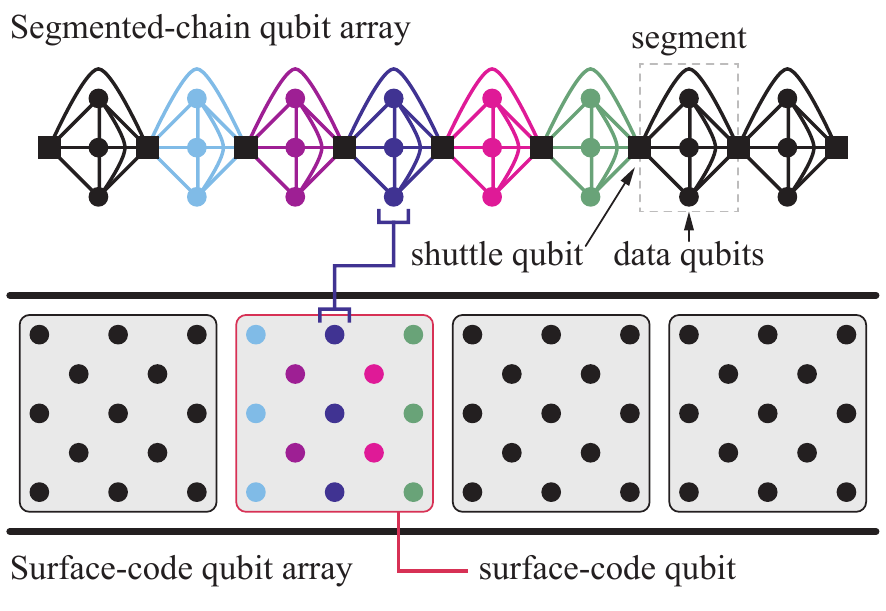}
\caption{
The surface code in a segmented chain qubit array. Each segment contains several data qubits (circles) and two shuttle qubits (squares). Qubits within the same segment are all-to-all connected, i.e.~two-qubit entangling gates (e.g.~CNOT gates) can be directly performed on any pair of qubits. Two nearest neighbouring segments are coupled by sharing one shuttle qubit. A surface-code logical qubit is encoded in several segments, and each segment provides qubits in one column of the surface code. CNOT gates can be performed on neighbouring surface-code qubits (see Sec.~\ref{sec:logical_gates}). These logical qubits form a one-dimensional quantum computer with nearest neighbouring interactions.
}
\label{fig:Q_chain}
\end{figure}

A surface-code logical qubit with the code distance $d$ is encoded in $d^2+(d-1)^2$ physical qubits, as shown in Fig.~\ref{fig:Q_chain}, in which the code distance is $d=3$. In a 2D array of physical qubits, implementing the surface code only requires interactions between neighbouring qubits~\cite{Fowler2009, Wang2011}.

In a 1D array of physical qubits with the segmented chain structure, we allocate one segment to each column of physical qubits in the square array (see Fig.~\ref{fig:Q_chain}): each physical qubit in the column maps to a data qubit in the segment; intra-column gates are performed using interactions within the segment; and gates between NN columns are realised via shuttle qubits.

The code distance of the surface code depends on the size of segments. We define the size of segments $s$ as the number of data qubits plus two shuttle qubits. For ion traps, this is the maximum number of ions in one trap; for qubit-resonator systems, this is the total number of qubits coupled to one resonator~(Fig.~\ref{fig:system}). In Fig.~\ref{fig:Q_chain}, the size of segments is $s = 5$. The number of qubits in a column is either $d$ or $d-1$ (long or short column). Limited by the segment size, the code distance $d \leq s-2$. In order to utilise the full computational power provided by the machine, we always choose $d = s-2$.

In each short column, one data qubit in the corresponding segment is not used in the surface-code encoding. These unused qubits are useful for suppressing logical memory errors and gating surface-code qubits, which will be discussed later.

In the following, we will show how to implement the surface code in the segmented chain qubit array.

\subsection{Stabiliser measurements}
\label{sec:circuits}

\begin{figure}[tbp]
\centering
\includegraphics[width=1\linewidth]{\figpath 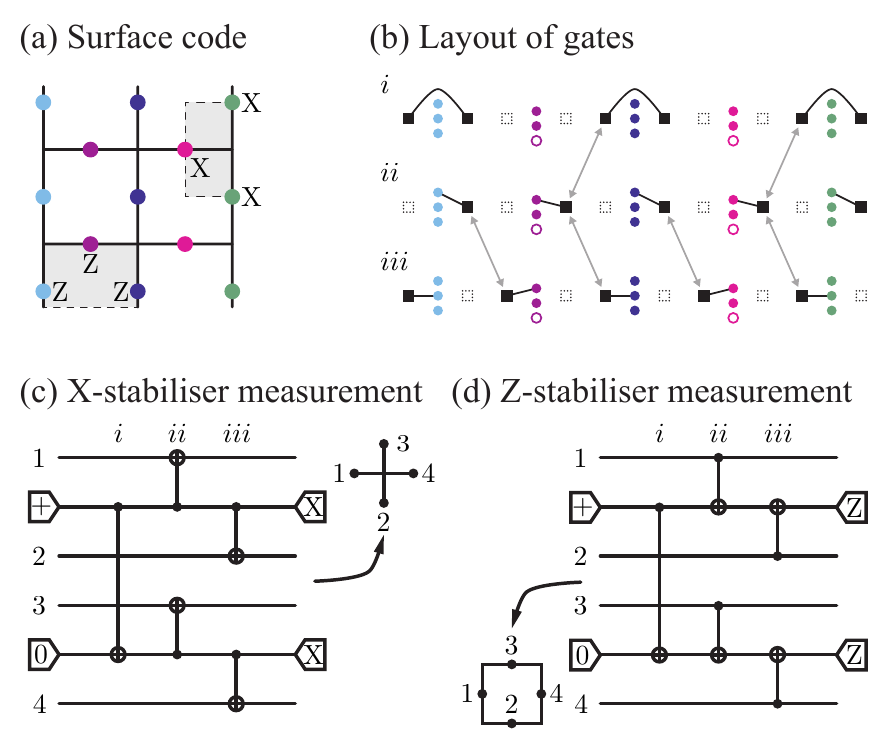}
\caption{
Surface-code stabiliser measurements in a segmented chain qubit array. (a) The lattice of the surface code. (b) The layout of two-qubit gates for measuring the first-row of X stabilisers. Circles are data qubits, and empty circles are unused data qubits in short columns. Squares are shuttle qubits, which are moved between nearest neighbouring columns as indicated by gray arrows. Black curves are CNOT gates. (c) Circuit of X-stabiliser measurements. (d) Circuit of Z-stabiliser measurements. Qubits 1, 2, 3 and 4 are data qubits, and other two qubits are shuttle qubits, which are initialised and measured in the circuit. The upper shuttle qubit is shared by segments of qubit-1 and qubit-2, and the lower shuttle qubit is shared by segments of qubit-3 and qubit-4. Qubit-2 and qubit-3 are in the same segment (column), which is a long column in (c) and a short column in (d).
}
\label{fig:circuit}
\end{figure}

The surface code is a stabiliser code, which means that the code can be defined using stabilisers~\cite{Nielsen2010}. As the same as many other stabiliser codes, the surface code has two sets of stabilisers (X and Z). Conventionally, stabilisers of the surface code are illustrated using the lattice in Fig.~\ref{fig:circuit}(a): each edge represents a physical qubit, each vertex represents an X stabiliser, and each plaquette represents a Z stabiliser~\cite{Dennis2002}. Typically a vertex is connected to four edges, and the corresponding X stabiliser is in the form $XXXX$, which is a tenser product of x-direction Pauli operators of four qubits on these edges. It is similar for a plaquette, whose perimeter is typically formed by four edges, and the corresponding Z stabiliser is in the form $ZZZZ$, which is a tenser product of z-direction Pauli operators of four qubits on these edges. Stabilisers on boundaries are special; each vertex (plaquette) on the left-right (top-bottom) boundaries only has three edges, i.e.~the corresponding stabiliser is a tenser product of three Pauli operators rather than four [see Fig.~\ref{fig:circuit}(a)].

Errors in the quantum computation are unexpected operations on the quantum state. The task of quantum error correction is to detect errors and undo them. In stabiliser codes, errors are detected by repeatedly measuring stabilisers. X and Z stabilisers are respectively used to detect phase and bit errors. For example if an error that is a phase-flip gate $Z$ occurs on a qubit, the sign of $X$ (the phase state) of the qubit is flipped, then the sign of the X stabiliser ($XXXX$ or $XXX$) containing this qubit is also flipped for once. Therefore, if we find that an X stabiliser is flipped, we can conclude that phase states of one or three qubits in this stabiliser have been flipped; if an X stabiliser is not flipped, then the phase states of either no qubit or two or four qubits have been flipped. It is similar for Z stabilisers. If a bit-flip error $X$ occurs, the sign of $Z$ (the bit state) of the qubit is flipped, and all corresponding Z stabilisers are flipped for once. An error changes either the phase state or the bit state or both them of a qubit, otherwise it is equivalent to an identity operation. We will discuss how to undo errors in Sec.~\ref{sec:error_rate}.

The protocol for stabiliser measurements in the segmented chain qubit array is shown in Fig.~\ref{fig:circuit}(b-d). Each column of physical qubits corresponds to a segment, and each segment has two shuttle qubits (Fig.~\ref{fig:Q_chain}). We call columns with $d$ qubits long columns and columns with $d-1$ qubits short columns. We use shuttle qubits as ancillaries to measure qubits as shown in circuits in Fig.~\ref{fig:circuit}(c,d). Stabilisers are measured row by row. In Fig.~\ref{fig:circuit}(b), we take the top row of X stabilisers (vertices) in Fig.~\ref{fig:circuit}(a) as an example. Firstly, shuttle qubits are initialised according to the circuit in Fig.~\ref{fig:circuit}(c). In step-$i$, two shuttle qubits of each long-column segment are entangled using interactions within the segment, corresponding to the step-$i$ CNOT gate in Fig.~\ref{fig:circuit}(c). In step-$ii$, shuttle qubits on the right side of long columns stay in long columns, shuttle qubits on the left side of long columns are {\it moved} leftward to short columns, and then each shuttle qubit interacts with a data qubit in the corresponding column to perform step-$ii$ CNOT gates in Fig.~\ref{fig:circuit}(c). In step-$iii$, all shuttle qubits are {\it moved} rightward and interact with data qubits to perform step-$iii$ CNOT gates in Fig.~\ref{fig:circuit}(c). Finally, shuttle qubits are measured to read stabiliser values according to the circuit in Fig.~\ref{fig:circuit}(c). We would like to remark that, in ion traps, shuttle qubits are physically moved between NN traps; but in qubit-resonator systems, moving shuttle qubits only means using different resonators. In five steps (including shuttle-qubit initialisation and measurement and three rounds of interactions), one row of X stabilisers are measured. By measuring X stabilisers row by row, we need $5(d-1)$ steps to complete the X-stabiliser measurements. Z-stabiliser measurements are similar [see Fig.~\ref{fig:circuit}(d)]. To measure Z stabilisers, the roles of long columns and short columns are exchanged, i.e.~two shuttle qubits of each short-column segment are entangled at the beginning, and then shuttle qubits are {\it moved} and interact with data qubits in the similar way. By measuring Z stabilisers row by row, we need $5d$ steps to complete Z-stabiliser measurements. Therefore, a full round of stabiliser measurements needs $5(2d-1)$ steps.

Circuits for measuring three-qubit stabilisers can be obtained from circuits in Fig.~\ref{fig:circuit}(c,d) by removing a data qubit and corresponding gates.

We can conclude that stabiliser measurements only require the initialisation and measurement operations on shuttle qubits and interactions involving shuttle qubits. Interactions between data qubits are not required. Single-qubit gates are not explicitly shown in the circuits, which may be needed for adjusting the initialisation/measurement basis. Similarly, single-qubit gates on data qubits are also not required. These operations required by stabiliser measurements form a set of universal operations, all operations on data qubits can be realised using these operations: we can transfer the state of a data qubit to a shuttle qubit to realise initialisation, measurement and single-qubit gates on the data qubit and transfer states of two data qubits to two shuttle qubits to realise two-qubit gates on data qubits. Realising data-qubit operations indirectly causes more errors, but because they are not used in stabiliser measurements, these extra errors will not change the performance of the quantum error correction significantly. In this paper, we assume that data qubits can be controlled directly, but the conclusion will be similar for systems without direct control on data qubits. As a remark, we have assumed that only one CNOT gate can be performed in a segment at a time.

\subsection{Post-correction error rate}
\label{sec:error_rate}

\begin{figure}[tbp]
\centering
\includegraphics[width=1\linewidth]{\figpath 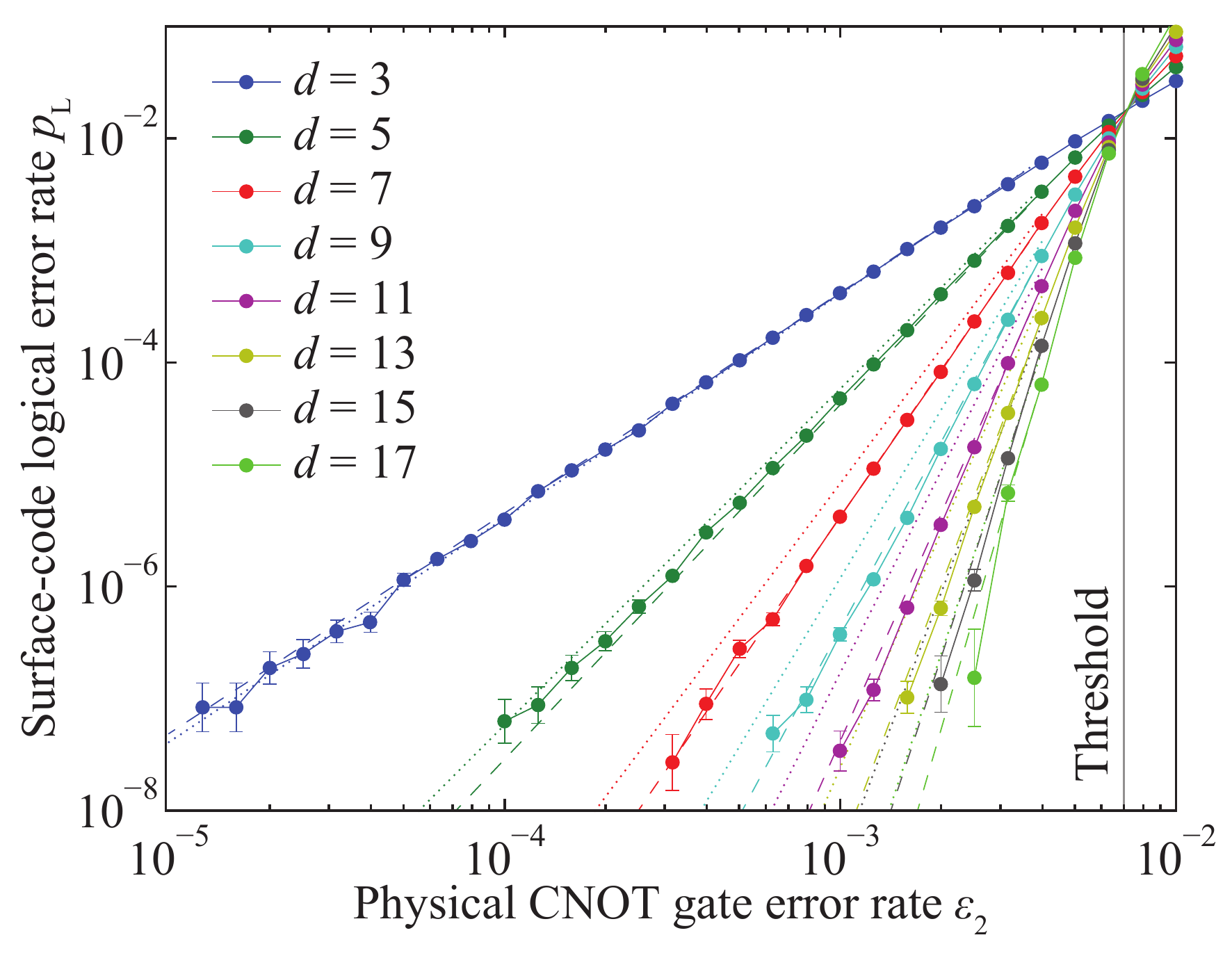}
\caption{
The rate of errors on a surface-code logical qubit per round of stabiliser measurements $p_{\rm L}$ as a function of the physical-qubit CNOT-gate error rate $\varepsilon_2$ and the code distance $d$. When the physical error rate is lower than the threshold marked by the vertical gray line, the logical error rate decreases with the code distance. Circles are data calculated numerically using the Monte Carlo method. Dashed lines are obtained by fitting circles (in the sub-threshold regime) using Eq.~(\ref{eq:scaling}). Dotted lines are calculated using Eq.~(\ref{eq:ef}). Error bars show one standard deviation, and error bars smaller than the size of circles have been removed from the figure.
}
\label{fig:surface_code}
\end{figure}

Stabiliser measurements provide us the information that we need for correcting errors. Stabilisers only reveal the parity of any errors on each stabiliser: if a stabiliser is flipped due to an odd number of errors, the event is called an error syndrome, which is an evidence of the presence of errors. We remark that if a stabiliser is not flipped, either there is no error or an even number of errors. Therefore, we need an algorithm to find out which qubits are affected by errors from error syndromes. Such an algorithm is called a decoder and is implemented using classical computers. Roughly speaking, the task of the decoder is to find out the most likely set of errors that can result in the given error syndromes (i.e.~stabiliser-measurement outcomes). Here, we outline the decoder used in this paper, and details can be found in Ref.~\cite{Wang2011}: the minimum-weight perfect matching algorithm is used to pair error syndromes~\cite{Kolmogorov2009}; in the lattice representing potential errors, diagonal edges are introduced to represent correlated errors; and the weight of an edge is determined by the rate of the corresponding error. Once errors are determined, we can undo errors by performing the inverse operation, i.e.~if we find that a phase-flip error $Z$ (a bit-flip error $X$) is on a qubit, the error can be corrected by performing a $Z$ ($X$) gate on the qubit; and we can record correction operations and only perform them until it is necessary, which is usually before a non-Clifford gate. Stabiliser-measurement outcomes themselves may be false due to errors in stabiliser measurement circuits, false outcomes can also be detected and corrected by the surface code~\cite{Wang2011}.

Two different sets of errors may result in the same error syndromes. Therefore, given error syndromes, correction operations selected by the decoder may be different from the actual errors to be corrected, resulting in some remaining errors on the state. These errors after the error correction cannot be detected by further stabiliser measurements, but frequently they are not harmful to the logical state. However, if the logical state is changed by a set of post-correction errors, the error correction has failed.

The rate of error correction failures, i.e.~the logical-qubit error rate, depends on the rate of errors in each operation on the physical-qubit level and the code distance. For the surface-code, usually when the physical-qubit error rate is lower than a threshold value, the logical-qubit error rate decreases with the code distance. We numerically studied the performance of the surface code using the Monte Carlo method. The error-rate threshold is determined; the logical-qubit error rate in the shallow sub-threshold regime (the physical error rate is relatively high and the code distance is small) is calculated directly, and the logical-qubit error rate in the deep sub-threshold regime is estimated using extrapolation.

We model the noise in the quantum computer as depolarising errors. Operations used in stabiliser measurements [Fig.~\ref{fig:circuit}(c,d)] include initialisations, measurements, Hadamard gates and CNOT gates (Hadamard gates are used for adjusting the initialisation/measurement basis). When a qubit is supposed to be initialised in the state $\ket{0}$, the qubit may be initialised in the incorrect state ($\ket{1}$) with the probability $\varepsilon_{\rm I}$. When a qubit is measured in the $0$/$1$ basis, the measurement outcome is incorrect with the probability $\varepsilon_{\rm M}$. Initialisations and measurements in the $+$/$-$ basis are realised using initialisations and measurements in the $0$/$1$ basis and Hadamard gates. A quantum gate with noise can be expressed as a superoperator $\mathcal{N}[U]$, where $[U]\rho = U\rho U^\dag$ represents the unitary gate, and $\mathcal{N}$ is a superoperator represents the noise. For single-qubit gates, the noise superoperator is
\begin{eqnarray}
\mathcal{N}_1 = (1-\frac{4}{3}\varepsilon_1)[\openone]
+ \frac{\varepsilon_1}{3}\sum_{a=0}^{3}[\SIG{(a)}{}].
\label{eq:N1}
\end{eqnarray}
For two-qubit gates, the noise superoperator is
\begin{eqnarray}
\mathcal{N}_2 = (1-\frac{16}{15}\varepsilon_2)[\openone]
+ \frac{\varepsilon_2}{15}\sum_{a=0}^{3}\sum_{b=0}^{3}[\SIG{(a)}{1}\SIG{(b)}{2}].
\end{eqnarray}
Here, $\varepsilon_1$ and $\varepsilon_2$ are rates of errors per gate, $\SIG{(a)}{i}$ is a Pauli operator of qubit-$i$, and $a = 0,1,2,3$ respectively correspond to $\openone$, $X$, $Y$ and $Z$. We assume that all these error rates are the same except the single-qubit error rate, which is assumed to be tenth of other error rates, i.e.~$\varepsilon_{\rm I} = \varepsilon_{\rm M} = 10\varepsilon_1 = \varepsilon_2$.

The noise in the identity operation (decoherence, memory errors) is modelled as the same as the noise in single-qubit gates, i.e.~the noise superoperator is the same as $\mathcal{N}_1$ [see Eq.~(\ref{eq:N1})]. We replace $\varepsilon_1$ with $\varepsilon_0$ to denote the rate of memory errors. The memory-error rate depends on the duration of the identity operation. We assume durations of initialisations, measurements and two-qubit gates are the same, and the duration of single-qubit gates is negligible compared with other operations. Then, we specify that $\varepsilon_0$ denotes the error rate during the time of a two-qubit gate. Implementing the surface code in the segmented chain qubit array, the time to perform one round of stabiliser measurements increases with the code distance. If we assume $\varepsilon_0$ is independent of the code distance, the error-rate threshold does not exist. Therefore, we assume that the rate of memory errors during one round of stabiliser measurement is equivalent to the error rate of two-qubit gates, i.e.~$\varepsilon_0$ depends on the code distance and $5(2d-1)\varepsilon_0 = \varepsilon_2$. We would like to remark that, in a segmented chain quantum computer, the scalability is achieved by using more segments in the chain instead of increasing the size of each segment. Therefore, the code distance $d$ and the required memory error rate are always finite and do not scale with the computer size. We will show that the required memory error rate is realistic for today's technologies, e.g.~in ion traps~\cite{Harty2014, Wang2017}.

\begin{table}[tbp]
\begin{center}
\begin{tabular}{|c|c||c|c|}
\hline
$\alpha$ & $0.5978$ & $\sigma_\alpha$ & $0.0058$  \\ \hline
$\beta$ & $2.9767$ & $\sigma_\beta$ & $0.0330$  \\ \hline
$\gamma$ & $-3.9819$ & $\sigma_\gamma$ & $0.0256$  \\ \hline
$\delta$ & $0.2923$ & $\sigma_\delta$ & $0.0413$  \\ \hline
\end{tabular}
\end{center}
\caption{
Parameters $\alpha$, $\beta$, $\gamma$ and $\delta$ and their standard deviations obtained by fitting data calculated using the Monte Carlo method, as shown in Fig.~\ref{fig:surface_code}.
}
\label{table}
\end{table}

The rate of errors on a surface-code qubit per round of stabiliser measurements $p_{\rm L}$ is plotted in Fig.~\ref{fig:surface_code}. The threshold of the CNOT gate error rate is at $\varepsilon^{\rm th}_2 \simeq 0.7\%$. If the physical error rate is lower than the threshold, the logical error rate decreases with the code distance. The logical error rate is fitted using the formula
\begin{eqnarray}
p_{\rm L} = \exp[(\alpha \log \varepsilon_2 + \beta) (d + \delta) + \gamma],
\label{eq:scaling}
\end{eqnarray}
where parameters $\alpha$, $\beta$, $\gamma$ and $\delta$ obtained from the fitting are given in Table~\ref{table}. As shown in Fig.~\ref{fig:surface_code}, this formula provides a good fit to the logical error rate in the shallow sub-threshold regime. In the following, we will use the same formula to estimate the logical error rate in the deep sub-threshold regime. An empirical formula~\cite{Fowler2012} used in the literature is
\begin{eqnarray}
p_{\rm L} = p_{\rm th} (\varepsilon_2/\varepsilon^{\rm th}_2)^{(d+1)/2},
\label{eq:ef}
\end{eqnarray}
where we take $p_{\rm th} = 0.02$. This empirical formula coincides with Eq.~(\ref{eq:scaling}) but parameters are different, i.e.~$\alpha = 0.5$, $\beta = -\alpha \log \varepsilon^{\rm th}_2 = 2.4809$, $\gamma = \log p_{\rm th} = -3.9120$ and $\delta = 0.5$. The logical error rate according to the empirical formula is also plotted in Fig.~\ref{fig:state_transfer}. We can find that the logical error rate estimated using the empirical formula is obviously higher than the value directly calculated using the Monte Carlo method, i.e.~Eq.~(\ref{eq:ef}) provides a conservative estimation of the logical error rate.

We would like to remark that the logical error rate in Fig.~\ref{fig:surface_code} is the rate of phase errors. Each X-stabiliser measurement for detecting phase errors needs two more Hadamard gates (for measuring shuttle qubits  in the $+$/$-$ basis) than a Z-stabiliser measurements [Fig.~\ref{fig:circuit}(c,d). Therefore, the chance that an X-stabiliser measurement reports a false outcome is higher than a Z-stabiliser measurement, and the rate of logical phase errors is slightly higher than the rate of logical bit errors. In our numerical simulations, we have considered the surface code in the orientation shown in Fig.~\ref{fig:circuit}(a), in which qubits in a column are in the same segment. As the dimension of the surface code in the vertical (column) direction is restricted, we can use more segments to increase the dimension in the horizontal direction. In this way, we can reduce the rate of logical bit errors exponentially as a function of the horizontal dimension (number of segments) at the price of increasing the rate of logical phase errors linearly. Therefore, it is essential to study logical phase error rather than logical bit errors.

Unused qubits in short-column segments can be used to reduce logical phase errors. Because of the all-to-all connectivity within each segment, exploiting these unused qubits we can change the surface code to a hybrid-boundary-condition code, in which the boundary condition along the horizontal direction is open (like the surface code) but the boundary condition along the vertical direction is closed (like the toric code). Because the toric code has a lower logical error rate due to the boundary condition, we can suppress logical phase errors in this way~\cite{Fowler2013}.

\subsection{Surface-code logical gates}
\label{sec:logical_gates}

\begin{figure*}[tbp]
\centering
\includegraphics[width=1\linewidth]{\figpath 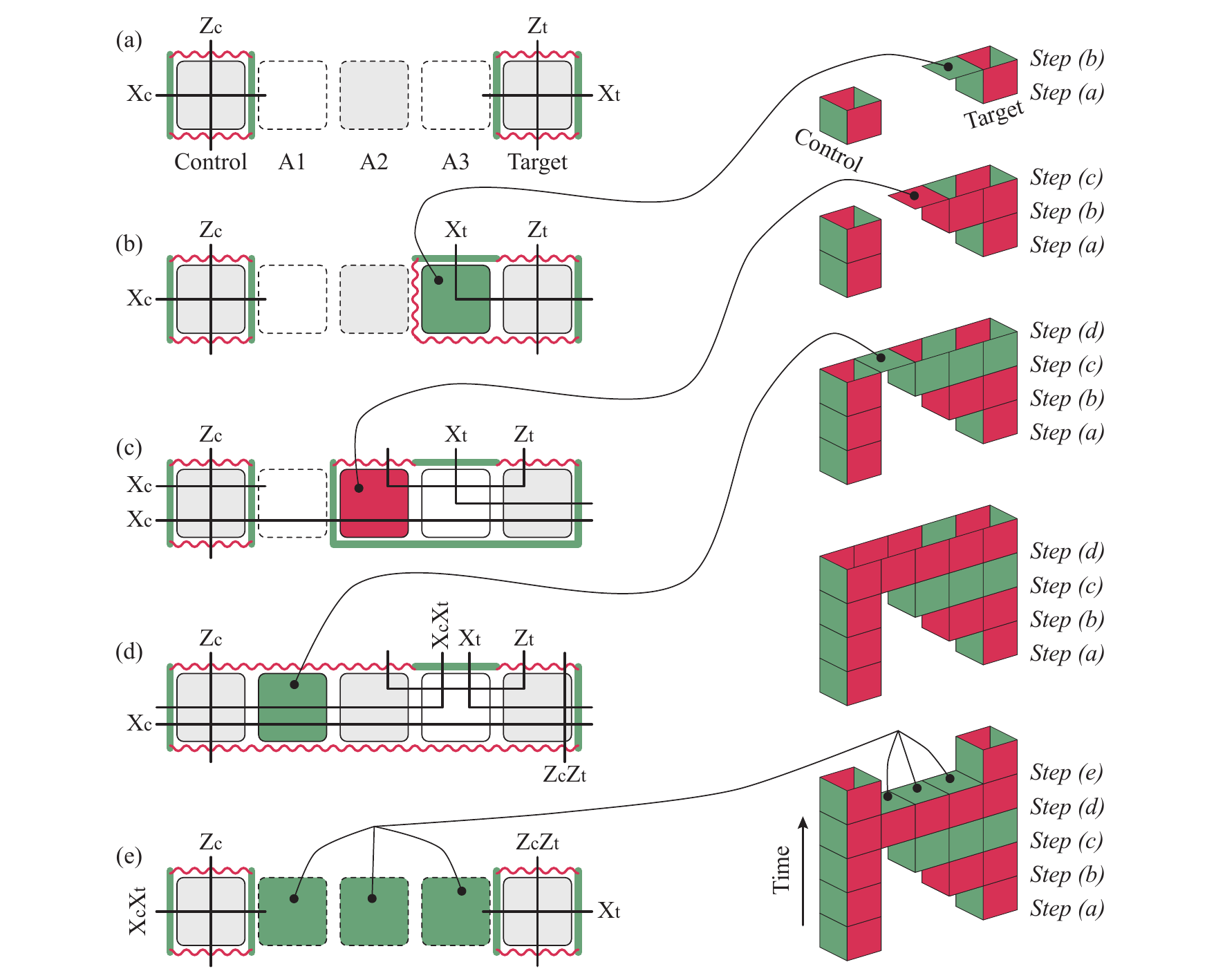}
\caption{
Protocol for CNOT gates in a one-dimensional array of surface-code logical qubits. In (a)-(e), squares denote surface-code qubits, red wave lines denote rough boundaries, and green bold lines denote smooth boundaries. A path (black line) connecting a pair of rough (smooth) boundaries denotes a logical $Z$ ($X$) operator. We use $X_{\rm c}$ and $Z_{\rm c}$ ($X_{\rm t}$ and $Z_{\rm t}$) to denote logical Pauli operators of the control (target) surface-code qubit. The protocol has four steps, and after each step stabilisers are measured for $h$ rounds, where $h \sim d$.
(a) To perform a CNOT gate on surface-code qubits Control and Target, we need three ancillary surface-code  qubits A1, A2 and A3. At the beginning, only Control and Target are used (as input).
(b) In the first step, data qubits in A3 are initialised in the state $\ket{+}$. Then $X_{\rm t}$ is deformed to a path connecting the top side of A3 and the right side of Target.
(c) In the second step, data qubits in A2 are initialised in the state $\ket{0}$. Then $Z_{\rm t}$ is deformed to a path connecting the top side of A2 and the top side of Target. Now, there is a long-path representation of $X_{\rm c}$, which is the path connecting the left side of Control and the right side of Target. Because the left side of A2 and the right side of Target belong to the same smooth boundary, a path connecting these two sides represents a logical identity.
(d) In the third step, data qubits in A1 are initialised in the state $\ket{+}$. The long-path representation of $X_{\rm c}$ is not changed in this step (but the gap at A1 is filled). Combining $Z_{\rm c}$ and $Z_{\rm t}$, a path connecting the top and bottom sides of Target represents $Z_{\rm c}Z_{\rm t}$. Similarly, combining $X_{\rm c}$ and $X_{\rm t}$, a path connecting the left side of Control and the top side of A3 represents $X_{\rm c}X_{\rm t}$.
(e) In the fourth step, data qubits in A1, A2 and A3 are measured in the $+$/$-$ basis. Then $X_{\rm c}X_{\rm t}$ is deformed to a path connecting the left and right sides of Control, and $X_{\rm t}$ is deformed to a path connecting the left and right sides of Target.
The right column illustrates the three-dimensional representation of the protocol, in which each block represents $h$ rounds of stabiliser measurements on a surface-code qubit. Red (green) surfaces correspond to rough (smooth) boundaries. The vertical direction represents the time. In (b), values of X stabilisers in A3 are determined, and the horizontal face of A3 is colored in green because it is equivalent to a smooth boundary. Similarly, the horizontal face of A2 is in red in (c), the horizontal face of A1 is in green in (d), and horizontal faces of A1, A2 and A3 are in green in (e).
}
\label{fig:blocks}
\end{figure*}

Fault-tolerant operations that can be directly performed on surface-code qubits include initialisations and measurements in the $0$/$1$ basis and the $+$/$-$ basis, Pauli gates, Hadamard gates and CNOT gates. A fault-tolerant initialisation (measurement) is realised by initialising (measuring) all data qubits in the corresponding basis, and a fault-tolerant Pauli gate is realised by performing a sequence of Pauli gates. In Sec.~\ref{appendix}, we give protocols of fault-tolerant Hadamard gates and CNOT gates in the 1D qubit array with the segmented chain structure.

Provided these fault-tolerant operations, the universality of quantum computing is completed by introducing magic states~\cite{Bravyi2005}. Magic states can be encoded using stabiliser measurements and distilled using fault-tolerant operations. High-fidelity (on the level of fault-tolerant operations) Clifford gates $S$ and non-Clifford gates $T$ can be realised using distilled magic states~\cite{Li2015}.

Our protocol for fault-tolerant CNOT gates only uses single-qubit operations and stabiliser measurements on a surface-code lattice with the width of one logical qubit (see the surface-code qubit array in Fig.~\ref{fig:Q_chain}), therefore it can be implemented in the segmented chain qubit array. CNOT gates are transverse gates of the surface code~\cite{Dennis2002}, however, transverse CNOT gates require interactions over the range of at least one logical qubit (i.e.~a qubit needs to talk to a qubit that is $\sim 2d^2$ qubits away in the 1D qubit array). Other protocols of fault-tolerant CNOT gates include braiding holes on a punched surface~\cite{Fowler2009} and lattice surgery~\cite{Horsman2012}, which only use neighbouring interactions but need the surface-code lattice to be two-dimensional with the minimum length of two logical qubits in both directions.

The overall flow of our protocol for fault-tolerant CNOT gates is shown in Fig.~\ref{fig:circuit}. Details of the protocol are given in Sec.~\ref{appendix}.

Each surface-code qubit is encoded in the lattice shown in Fig.~\ref{fig:circuit}(a). We call top and bottom boundaries rough boundaries (formed by edges with open ends), and left and right boundaries smooth boundaries (formed by edges without open ends). For any path formed by a sequence of edges connecting two rough boundaries, if we perform the $Z$ gate on each qubit on the path, the combined gate is equivalent to a $Z$ gate on the logical qubit. Similarly, a sequence of $X$ gates connecting two smooth boundaries is equivalent to a $X$ gate on the logical qubit. Understanding this property is important for understanding our protocol for the fault-tolerant CNOT gate shown in Fig.~\ref{fig:blocks}(a-e). In the following, we use $X_{\rm c}$ and $Z_{\rm c}$ ($X_{\rm t}$ and $Z_{\rm t}$) to denote logical Pauli gates, i.e.~operators, of the control (target) surface-code qubit.

In each step of the protocol, some paths representing logical Pauli gates are deformed. For example, in the first step [Fig.~\ref{fig:blocks}(b)], data qubits in A3 are initialised in the state $\ket{+}$, which is an eigenstate of $X$ with the eigenvalue $+1$. Therefore, $X_{\rm t}$ extends to the A3 region from the Target region, because $X$ gates in the A3 region do not have any effect the state. It is similar for the second and third steps. For the fourth step [Fig.~\ref{fig:blocks}(e)], because data qubits are measured in the $+$/$-$ basis in A1, A2 and A3, $X$ gates in these regions also do not have any effect on the state (up to a known phase). Therefore, $X_{\rm c}X_{\rm t}$ and $X_{\rm t}$ shrunk back to Control and Target regions, respectively. By deforming logical Pauli gates, we realise a transformation from $Z_{\rm c}$, $Z_{\rm t}$, $X_{\rm c}$ and $X_{\rm t}$ to $Z_{\rm c}$, $Z_{\rm c}Z_{\rm t}$, $X_{\rm c}X_{\rm t}$ and $X_{\rm t}$, respectively, which is a CNOT gate.

As shown in Fig.~\ref{fig:blocks}(b-d), in the second step, bottom sides of A3 and Target are changed from rough-boundary to smooth-boundary; and in the third step, bottom sides of A2, A3 and Target are changed back from smooth-boundary to rough-boundary. Changing the property of the boundary is realised by extending or shrinking the lattice for half of the length of an edge [see the lattice in Fig.~\ref{fig:circuit}(a)], in which we need to use unused data qubits in short-column segments (see Sec.~\ref{appendix} for details).

The three-dimensional illustration of the protocol is shown in Fig.~\ref{fig:blocks}(f-j). Each block has the dimension $\sim d\times d\times d$ and represents $\sim d$ rounds of stabiliser measurements on a surface-code qubit. Each fault-tolerant CNOT gate has in total $14$ blocks. There are $16$ blocks in Fig.~\ref{fig:blocks}(j), but two of them are due to two input surface-code qubits. One can find that the distance between any pair of disconnected red (green) strips is $\sim d$. Because red and green strips represent rough and smooth boundaries respectively, the distance between strips corresponds to the code distance, i.e.~the minimum number of single-qubit errors that can change the logical state but cannot be detected by stabilisers. Because the strip distance is $\sim d$, our protocol is fault-tolerant.

\section{One-dimensional four-qubit gauge code}
\label{sec:gauge}

\begin{figure}[tbp]
\centering
\includegraphics[width=1\linewidth]{\figpath 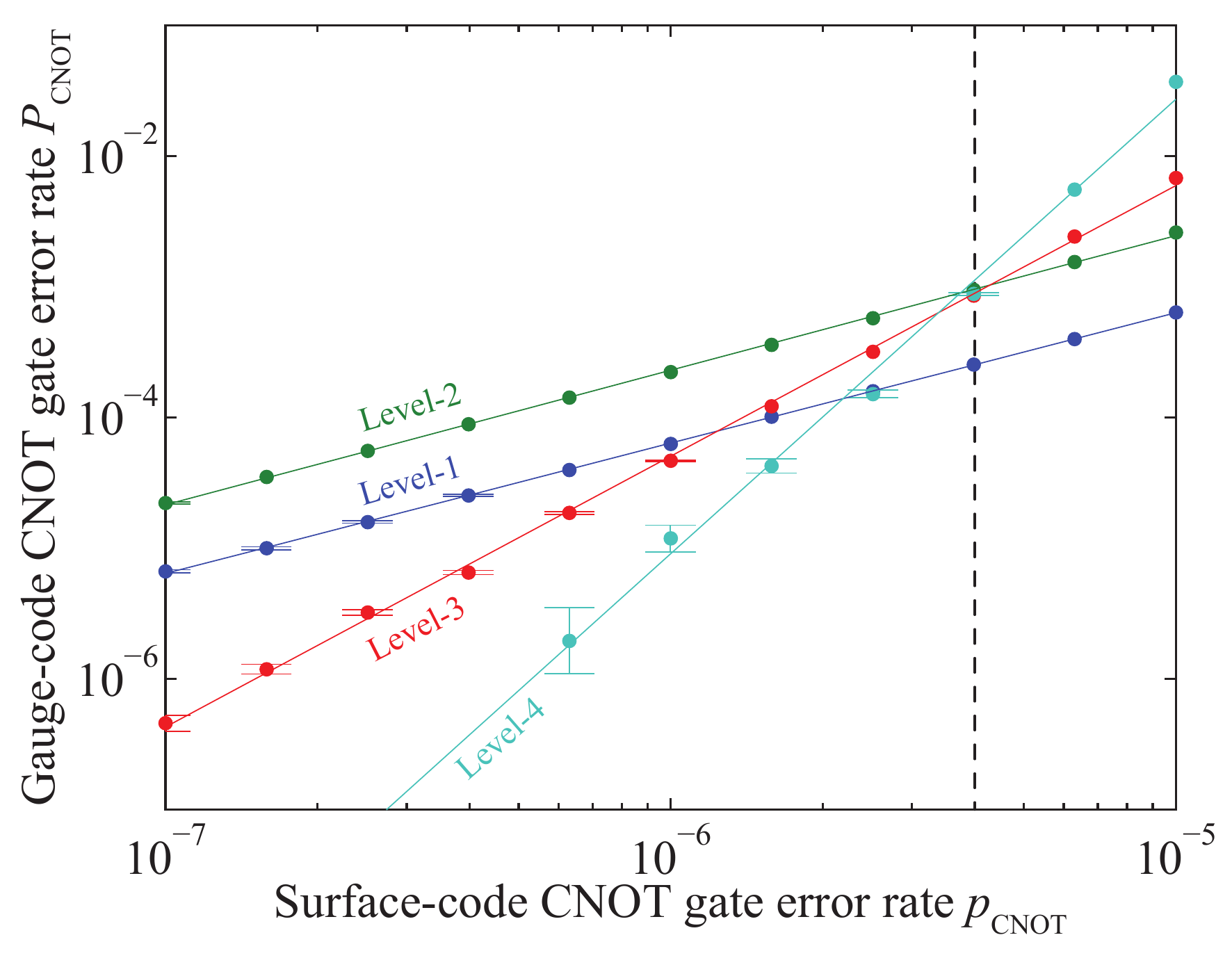}
\caption{
The logical CNOT gate error rate of the four-qubit gauge code $P_{\rm CNOT}$ as a function of the logical CNOT gate error rate of the surface code $p_{\rm CNOT}$. Gauge-code logical qubits are encoded in surface-code logical qubits, and the gauge code is concatenated. The level of the gauge-code concatenation is marked in the figure. The surface-code CNOT gate error rate is $p_{\rm CNOT} = 14dp_{\rm L}$. Error bars show one standard deviation, and error bars with invisible gaps have been removed from the figure.
}
\label{fig:gauge4}
\end{figure}

In the segmented chain qubit array, the code distance of the surface code is limited by the size of segments. If the logical error rate provided by the surface code is not low enough for implementing a quantum algorithm, we need another code on top of the surface code to further reduce the logical error rate.

The array of surface-code qubits is 1D and only has NN interactions (Fig.~\ref{fig:Q_chain}). As shown in Fig.~\ref{fig:blocks}, to perform the CNOT gate on a pair of surface-code qubits, we need three surface-code qubits between them as ancillaries. In the surface-code qubit array, we can choose one surface-code qubit to carry the information in every four of them, and other three surface-code qubits are used as ancillary qubits for performing CNOT gates between information qubits. In this way, we need four surface-code qubits to actually encode one bit of information. We can more efficiently use surface-code qubits by removing some ancillary qubits. The state of a surface-code qubit can be transferred to the NN surface-code qubit (see Sec.~\ref{appendix}). Therefore, in the extreme case that we have only three ancillary surface-code qubits at all, a CNOT gate can be performed by moving these three ancillary qubits to the right place. However, in this case, CNOT gates cannot be performed in parallel. In the following, we assume that only one in four surface-code qubits is the information qubit, so that CNOT gates can be performed in parallel between NN information surface-code qubits.

The quantum error correction in a 1D qubit array with NN interactions has been studied in the literature. The four-qubit gauge code is a successful code for 1D quantum error correction, whose threshold is estimated to be about $10^{-5}$~\cite{Stephens2009}. In this paper, we will focus on this four-qubit gauge code.

To study the performance of the four-qubit gauge code implemented using surface-code qubits, we need to know the error rate of surface-code logical operations. Operations required by the four-qubit gauge code includes initialisations and measurements in the $0$/$1$ basis and the $+$/$-$ basis, CNOT gates and SWAP gates between NN qubits. We assume that surface-code stabilisers are measured for $d$ rounds after a surface-code qubit is initialised or before measured. Therefore, we estimate the error rate of initialisations and measurements on surface-code qubits as $p_{\rm I/M} = dp_{\rm L}$. We assume that surface-code stabilisers are also measured for $h = d$ rounds after each step in the surface-code CNOT gate (Fig.~\ref{fig:blocks}). Therefore, we estimate the logical error rate of surface-code CNOT gates as $p_{\rm CNOT} = 14dp_{\rm L}$, where $14$ is number of blocks in the surface-code CNOT gate, and each block corresponds to $d$ rounds of stabiliser measurements on a surface-code qubit. A SWAP gate is realised by three CNOT gates, and its error rate is $p_{\rm SWAP} = 3p_{\rm CNOT}$. The rate of logical memory errors depends on the duration of the logical identity operation. For the duration of initialisations and measurements, the rate of memory errors is $p_0 = dp_{\rm L}$; for CNOT gates, it is $4p_0$; for SWAP gates, it is $12p_0$. We remark that these surface-code logical error rates are only for phase-flip errors, and it is similar for bit-flip errors.

The method of estimating the error rate of logical operations used here, which is calculating the space-time volume of stabiliser measurements~\cite{Fowler2012}, is not strictly accurate. However, a direct calculation of the logical error rate using the Monte Carlo method, e.g.~for the CNOT gate, requires a simulation of four logical qubits for $\sim 5d$ rounds of stabiliser measurement, which would be much harder than the numerical calculation that we have done in this paper (which used about 160,000 CPU hours). We have assumed that errors in logical CNOT gates are depolarised for simplification, which can also cause inaccuracy. All these assumptions in our numerical simulations will only change our result of the segment size slightly, because the logical error rate changes rapidly with the code distance. According to Eq.~(\ref{eq:ef}), by increasing the segment size by two qubits, the logical error rate can be reduced by a factor of $7$ ($70$) for the physical error rate $0.1\%$ ($0.01\%$).

Based on the our estimation of surface-code logical error rates, the error rate of gauge-code logical qubits is calculated using the Monte Carlo method, and the result is plotted in Fig.~\ref{fig:gauge4}. The code distance of the concatenated four-qubit gauge code is $2^n$, where $n$ is the level of concatenation. For the first-level concatenation, the code can only detect errors, because the code distance is $2$. From the second-level concatenation, the code starts to have the ability of correcting errors. In Fig.~\ref{fig:gauge4}, for concatenation levels $n=2,3,4$, a crossing point at $p_{\rm CNOT} = 4\times 10^{-6}$ is observed, which indicates a threshold: if the surface-code logical error rate is lower than the threshold, the gauge-code logical error rate can be reduced by increasing the level of concatenation.

This threshold of the 1D quantum error correction is lower than the threshold $10^{-5}$ reported in Ref.~\cite{Stephens2009}, because a different model of the noise is used. A recent paper proposed a protocol for the 1D quantum error correction using concatenated two-qubit repetition code~\cite{Jones2016}, in which an error-rate crossing at $\sim 10^{-4}$ is observed between and error-correction concatenation level and error-detection concatenation levels. This crossing may indicate a threshold higher than the four-qubit gauge code. In our protocol, the code on top of the surface code can be any code that only uses NN interactions in a 1D qubit array.

\begin{table}[tbp]
\begin{center}
\begin{tabular}{|c|c|c|c|c|}
\hline
$n$ & $\kappa_n$ & $\eta_n$ & $\sigma_{\kappa_n}$ & $\sigma_{\eta_n}$  \\ \hline \hline
$1$ & $0.9973$ & $4.1141$ & $0.0463$ & $0.0463$  \\ \hline
$2$ & $1.0303$ & $5.8552$ & $0.0815$ & $0.0815$  \\ \hline
$3$ & $2.0717$ & $18.7274$ & $0.2723$ & $0.2723$  \\ \hline
$4$ & $3.4795$ & $36.4548$ & $1.4622$ & $1.4622$  \\ \hline
\end{tabular}
\end{center}
\caption{
Parameters $\kappa_n$ and $\eta_n$ and their standard deviations obtained by fitting data calculated using the Monte Carlo method, as shown in Fig.~\ref{fig:gauge4}.
}
\label{GCtable}
\end{table}

The error rate $P_{\rm CNOT}$ of CNOT gates on gauge-code logical qubits increases with the surface-code error rate $p_{\rm CNOT}$. This dependence can be described using the formula
\begin{eqnarray}
P_{\rm CNOT} = \exp(\kappa_n \log p_{\rm CNOT} + \eta_n),
\label{eq:GCscaling}
\end{eqnarray}
where parameters $\kappa_n$ and $\eta_n$ depend on the concatenation level $n$. By fitting data in Fig~\ref{fig:gauge4}, we obtain parameters $\kappa_n$ and $\eta_n$ as given in Table~\ref{GCtable}. For first two levels of concatenations, $\kappa_1,\kappa_2 \sim 1$, which implies that the error correction does not work ($P_{\rm CNOT} \propto p_{\rm CNOT}^{\kappa_n}$). The first-level concatenation does not work because it can only detect errors. The second-level concatenation does not work because of two-qubit errors that cannot be corrected by the code with distance $4$. Therefore, the third level is the minimum level of encoding in order to take the advantage of the four-qubit gauge code to reduce errors.

\section{Fault-tolerant quantum computing}
\label{sec:FTQC}

\begin{figure}[tbp]
\centering
\includegraphics[width=1\linewidth]{\figpath 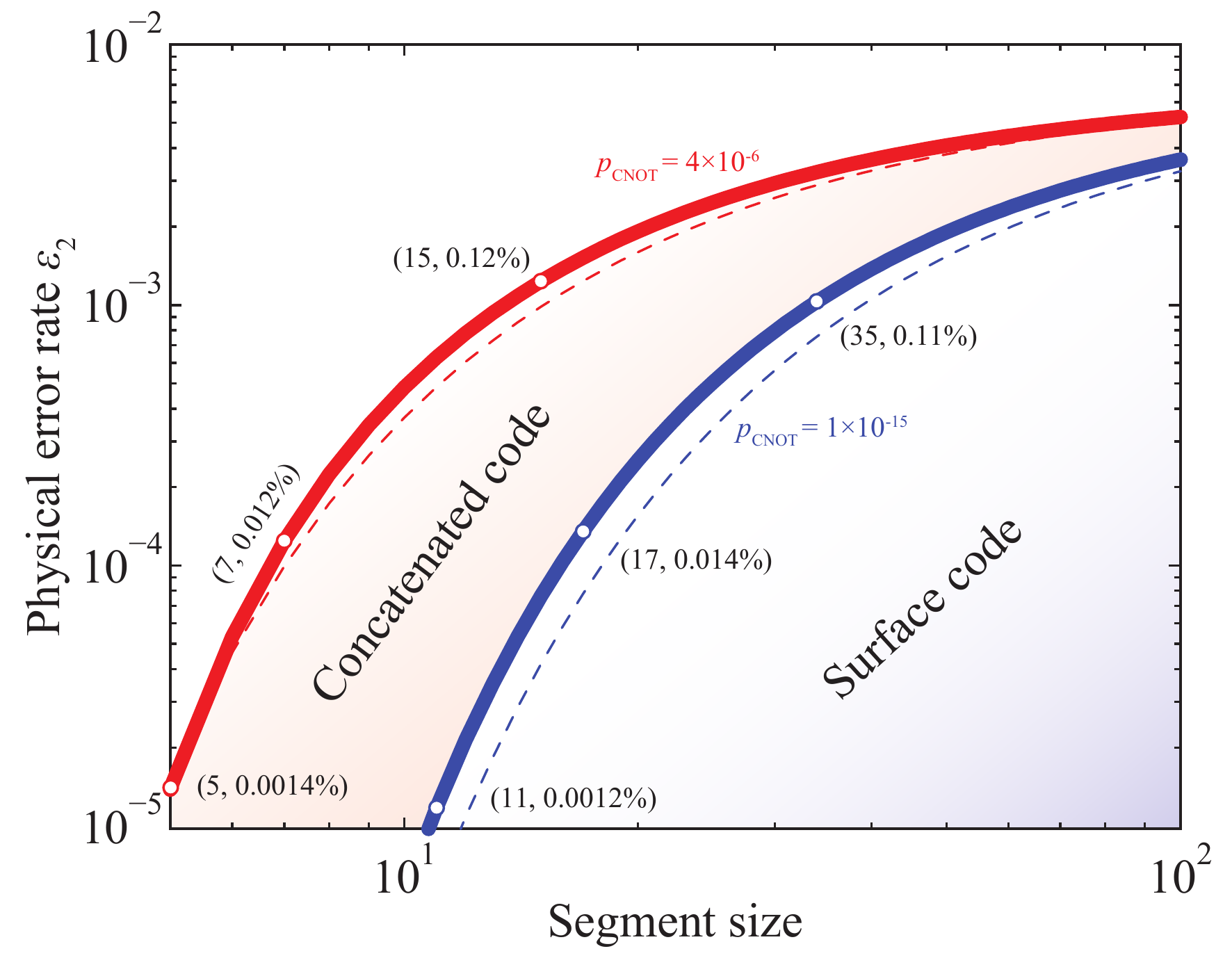}
\caption{
The physical error rate $\varepsilon_2$ and the segment size $s$ required to achieve the surface-code logical error rate $p_{\rm CNOT}$. The red curve corresponds to $p_{\rm CNOT} = 4\times 10^{-6}$, and the blue curve corresponds to $p_{\rm CNOT} = 10^{-15}$. These two curves are calculated using Eq.~(\ref{eq:scaling}) with parameters given in Table~\ref{table}. Results obtained using Eq.~(\ref{eq:ef}) are also plotted in the figure as dashed curves. Errors are firstly corrected using the surface code. When the physical error rate is not low enough or segments are not large enough (e.g.~to achieve $p_{\rm CNOT} = 10^{-15}$), we need the concatenated code on top of the surface code to further correct errors. $p_{\rm CNOT} = 4\times 10^{-6}$ is the threshold of the regime that the concatenated code works, therefore the red curve is the threshold of the overall protocol.
}
\label{fig:surface_code_logical_error_rate}
\end{figure}

In the segmented chain qubit array, the overall protocol for fault-tolerant quantum computing depends on the logical error rate required by the computing task, the rate of physical errors and the size of segments. If the surface code cannot suppress the logical error rate to the level required by the task, we need to use the concatenated four-qubit gauge code to further reduce the logical error rate. In order to use the concatenated code, the surface code has to firstly suppress the logical error rate to be lower than $p_{\rm CNOT} = 4\times 10^{-6}$, which leads to a threshold of the physical error rate. In Fig.~\ref{fig:surface_code_logical_error_rate}, this threshold of the physical error rate is plotted as a function of the segment size. If the physical error rate is $\varepsilon_2 = 0.12\%$, we need segments with more than $15$ qubits to build a fault-tolerant quantum computer. If the physical error rate can be reduced to $\varepsilon_2 = 0.012\%$, the minimum size of segments can be reduced to $7$.

When each segment is large enough, the surface code itself is enough for many quantum-computing tasks. As shown in Fig.~\ref{fig:surface_code_logical_error_rate}. Given the physical error rate $\varepsilon_2 = 0.11\%$ and segments with about $35$ qubits in each one of them, the surface-code CNOT gate error rate is $p_{\rm CNOT} \simeq 10^{-15}$, which is enough for implementing the Shor's algorithm with a thousand qubits~\cite{Fowler2012, Joe2016}. Similarly, if the physical error rate can be reduced to $\varepsilon_2 = 0.014\%$, the segment size only needs to be $17$ to achieve the same logical error rate.

\begin{figure}[tbp]
\centering
\includegraphics[width=1\linewidth]{\figpath 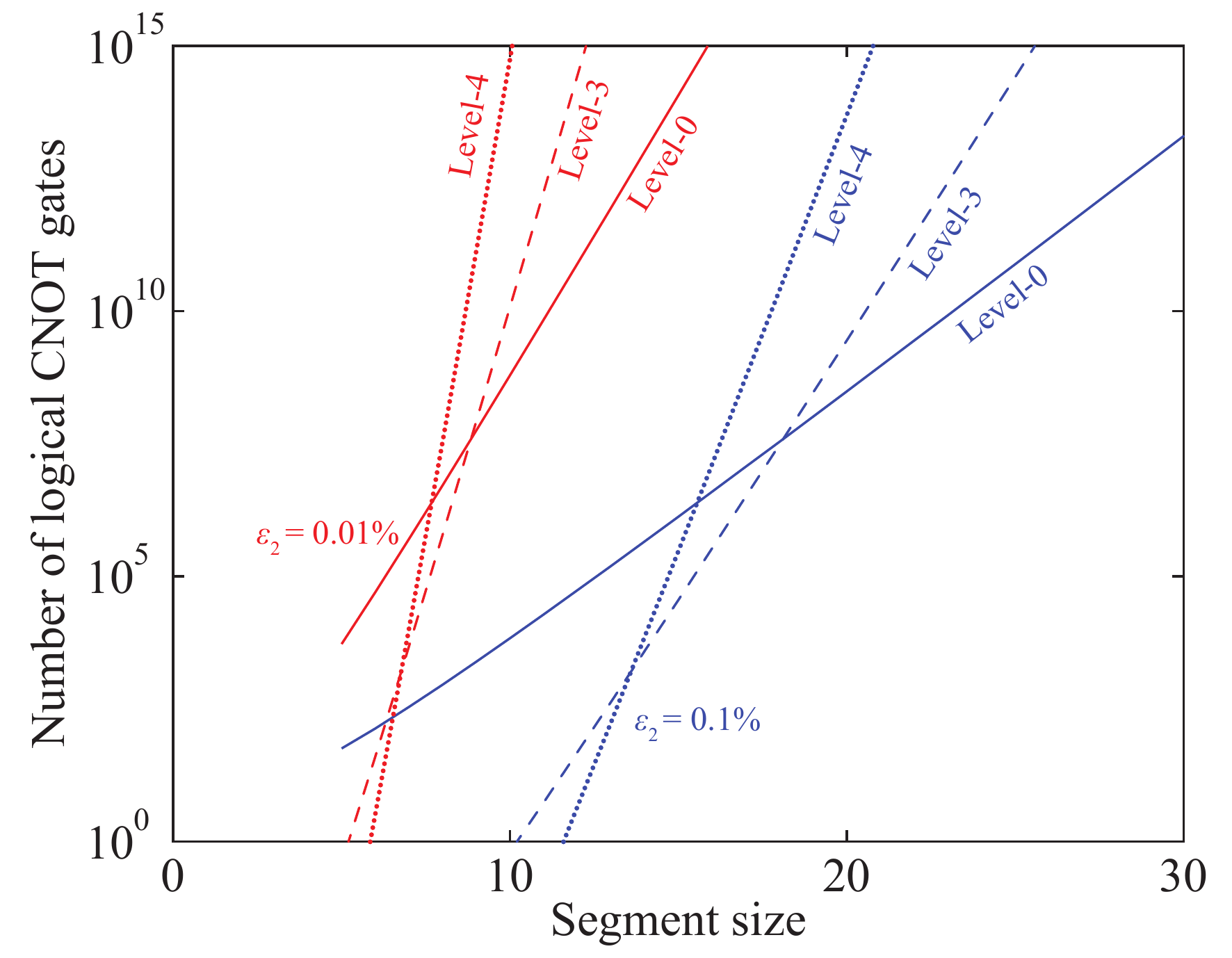}
\caption{
The average number of logical CNOT gates that can be performed in a quantum algorithm before getting one logical error. For the level-0 encoding, only the surface code is used to correct errors, and the number of gates is $1/p_{\rm CNOT}$. For the level-3 and level-4 encoding, the number of gates is $1/P_{\rm CNOT}$. $p_{\rm CNOT}$ is calculated using Eq.~(\ref{eq:scaling}), and $P_{\rm CNOT}$ is calculated using Eq.~(\ref{eq:GCscaling}). Parameters in these two equations are given in Table~\ref{table} and Table~\ref{GCtable}.
}
\label{fig:gauge4_logical_error_rate}
\end{figure}

The performance of the overall protocol is plotted in Fig.~\ref{fig:gauge4_logical_error_rate}. The level-0 encoding means that only the surface code is used to correct errors. The performance of the surface-code-only error correction, the third-level and fourth-level concatenated gauge codes are compared. Given the gate error rate $\varepsilon_2 = 0.1\%$ and using the fourth level concatenation, $10^{15}$ logical CNOT gates can be achieved with $21$ qubits in each segment. The concatenation is expensive. Using the four-qubit gauge code, to encode a higher-level qubit, we need six lower-level qubits. If we use only one in every four surface-code qubits as the information qubit, each logical qubit with the level-$n$ concatenation requires $4\times 6^n$ surface-code qubits, i.e.~the third-level (fourth-level) encoding needs $864$ ($5184$) surface-code qubits per gauge-code logical qubit. As shown in Fig.~\ref{fig:gauge4_logical_error_rate}, using the concatenated code can reduce the required segment size but the effect is modest especially when the physical error rate is as low as $\varepsilon_2 = 0.01\%$.

\section{Conclusions}
\label{sec:Conclusions}

We have discussed fault-tolerant quantum computing in 1D quantum computers with the segmented chain structure. Given the state-of-the-art error rate $0.1\%$, the size of each segment must be at least $15$ qubits for fault-tolerance to be of benefit using the surface code concatenated with the 1D gauge code, and $35$ qubits for large scale algorithms such as Shor's algorithm to be implemented only using the surface code. Each segment is a small quantum processor with all-to-all connections among qubits. Segments with $4$ or $5$ qubits have been demonstrated with ion traps~\cite{Choi2014, Debnath2016} and superconducting qubits~\cite{Takita2016}, and the qubit number in each segment in these platforms can be extended to tens or even more qubits~\cite{Choi2014, Islam2013, Jurcevic2014, Kakuyanagi2016}. The disadvantage of the segmented chain structure is the computing speed. Because the all-to-all connectivity within each segment is due to the coupling to the same phonon or photon modes, interactions between qubits in the same segment could not be switched on simultaneously. As a result, segmented chain 1D quantum computers need more operation cycles than 2D quantum computers by a factor determined by the segment size. Therefore, a longer coherence time is required. In ion traps, the coherence time of qubits is about $50$ sec~\cite{Harty2014}, which is $500,000$ times longer than $\sim 100$ $\mu$s the time cost of a two-qubit gate~\cite{Lucas, Takita2016}, i.e.~the memory error rate $\varepsilon_0 \sim 2\times 10^{-6}$, which allows a segment size as large as about $50\sim \varepsilon_2/(10\varepsilon_0)$ qubits for the gate error rate $\varepsilon_2 = 0.1\%$. This coherence time of ion qubits can be increased by a factor of $12$ by using dynamical decoupling~\cite{Wang2017}. Such a ratio of coherence time to gate time is still a challenge for superconducting qubits~\cite{Devoret2013}. Some recent works have been focusing on increasing the coherence time of superconducting qubits~\cite{Gustavsson2016} or coupling them to quantum memories, e.g. nitrogen-vacancy centres in diamond~\cite{Saito2013}. We remark that the computing speed also depends on the time cost of each operation cycle. As an alternative approach of building a quantum computer, the segmented chain structure avoids the need to expand the qubit array to higher dimensions, which reduces the complexity of the quantum computer and allows us to design the quantum computer based on the well-developed 1D quantum technologies and on-chip integrated circuit manufacturing technologies.

\begin{acknowledgments}
This work was supported by the EPSRC National Quantum Technology Hub in Networked Quantum Information Technology (EP/M013243/1). The authors would like to acknowledge the use of the University of Oxford Advanced Research Computing (ARC) facility in carrying out this work. http://dx.doi.org/10.5281/zenodo.22558.
\end{acknowledgments}

\appendix

\section*{Appendix}
\label{appendix}

\begin{figure*}[tbp]
\centering
\includegraphics[width=1\linewidth]{\figpath 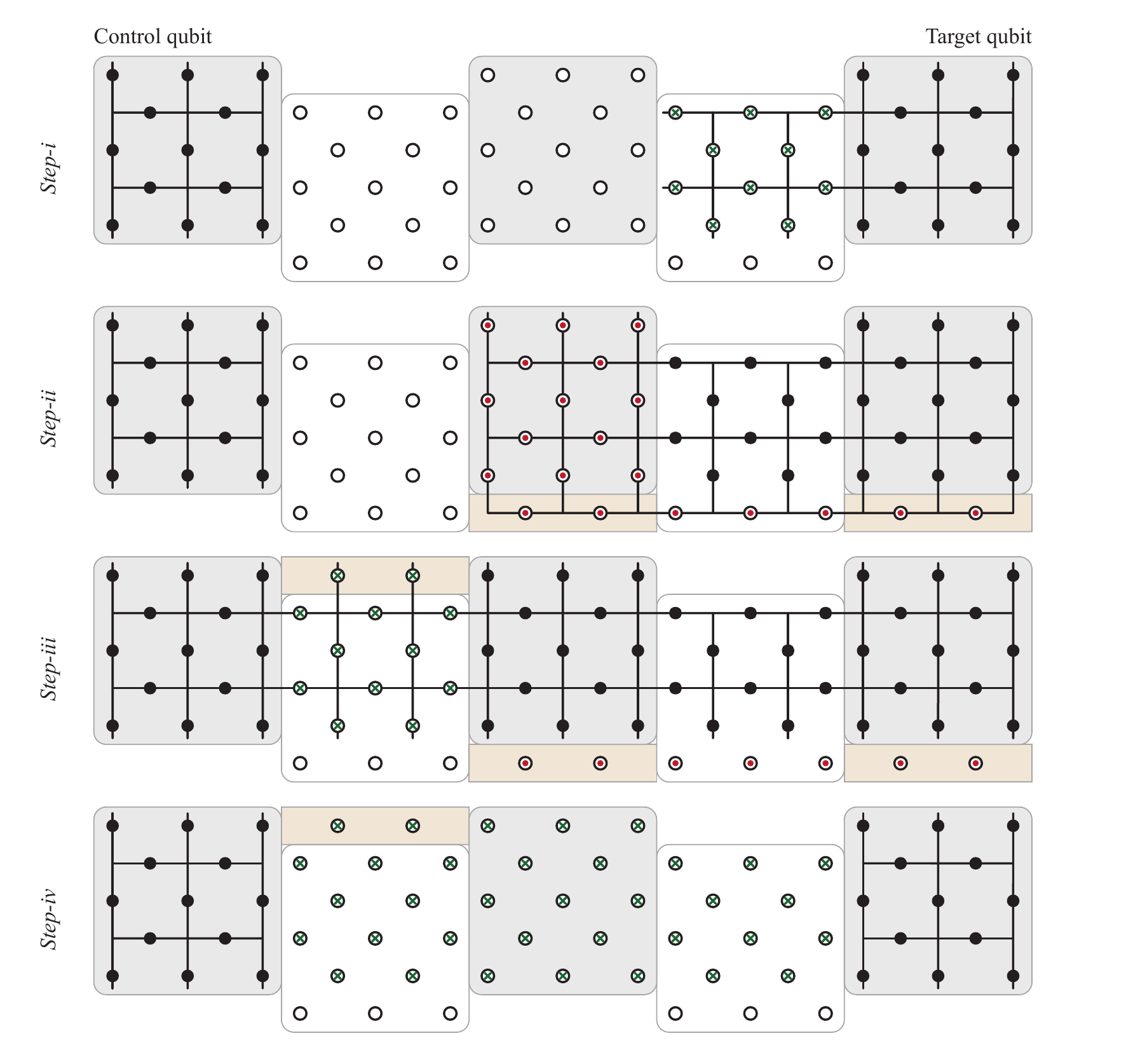}
\caption{
Protocol for CNOT gates on surface-code logical qubits. After each step, stabilisers are measured for $h\sim d$ rounds. Each circle denotes a data qubit. For each step qubits are divided into four groups: on-lattice solid circles, on-lattice circles with marks (green crosses or red dots), off-lattice circles with marks and off-lattice empty circles. In each step, on-lattice circles with green crosses (red dots) are initialised in the state $\ket{+}$ ($\ket{0}$), and off-lattice circles with green crosses (red dots) are measured in the $+$/$-$ ($0$/$1$) basis. Stabiliser measurements after the step are performed according to the lattice: each vertex corresponds to an X stabiliser, and each plaquette corresponds to a Z stabiliser. Solid circles are involved in both stabiliser measurements before and after the step, on-lattice (off-lattice) circles with marks are only involved in stabiliser measurements after (before) the step, and empty circles are involved in stabiliser measurements neither before nor after the step. Each square panel denotes a surface-code qubit. Qubits out of panels (in rectangle bars) are unused qubits of short columns.
}
\label{fig:CNOT}
\end{figure*}

\begin{figure*}[tbp]
\centering
\includegraphics[width=1\linewidth]{\figpath 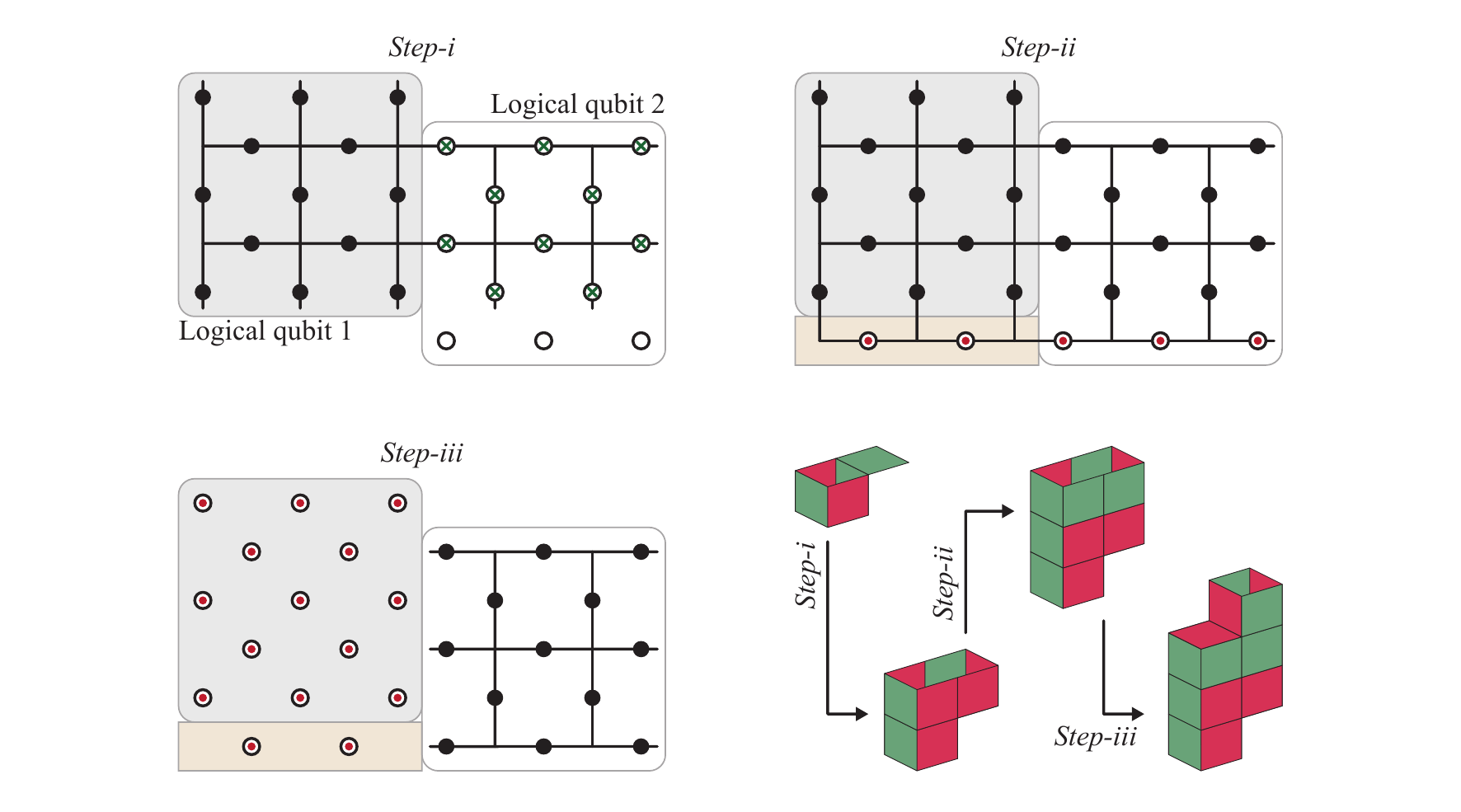}
\caption{
Protocol for Hadamard gates on surface-code logical qubits and its three-dimensional representation. After each step, stabilisers are measured for $h\sim d$ rounds. The notions are the same as explained in the caption of Fig.~\ref{fig:CNOT}. By implementing these three steps in the figure, the state of the logical qubit 1 is transferred to the logical qubit 2, and the orientation of the lattice is rotated. To complete the logical Hadamard gate, we also need to perform the Hadamard gate on each data qubit of the logical qubit 2, then its lattice is rotated to be the same as the lattice of the logical qubit 1, and the logical state is transformed according to the Hadamard gate. The overall operation is a state transfer from the logical qubit 1 to the logical qubit 2 followed by a logical Hadamard gate. The state of the logical qubit 2 can be transferred back to the logical qubit 1 without rotating the lattice as shown in Fig.~\ref{fig:state_transfer}. In the three-dimensional representation, the distance between any pair of disconnected red (green) strips is $\sim d$, therefore this protocol is fault-tolerant.
}
\label{fig:Hadamard}
\end{figure*}

\begin{figure*}[tbp]
\centering
\includegraphics[width=1\linewidth]{\figpath 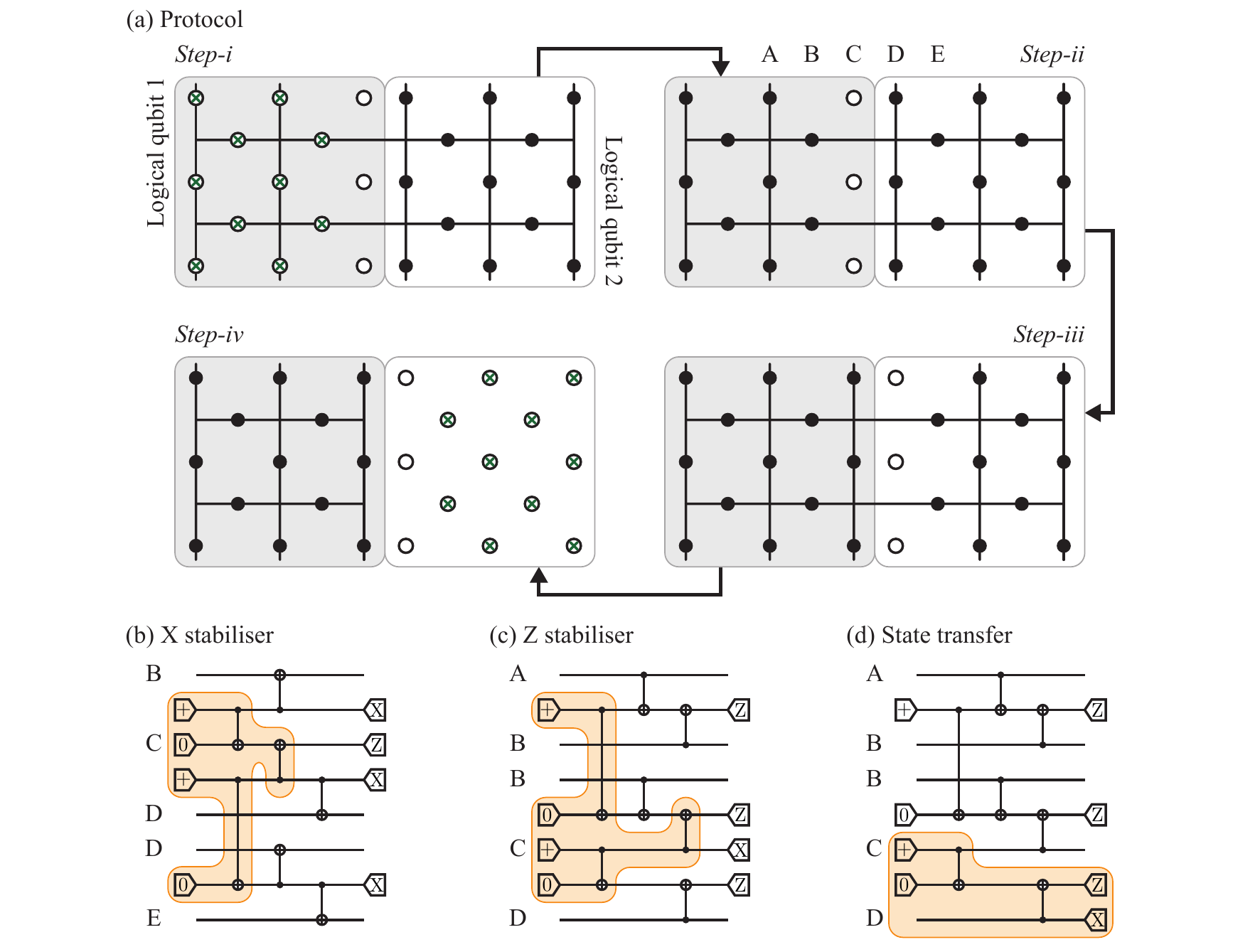}
\caption{
Protocol for the state transfer between surface-code logical qubits. The state transfer from the logical qubit 2 to the logical qubit 1 is achieved by extending the lattice from the logical qubit 2 to the logical qubit 1 and then shirking the lattice to the logical qubit 1. (a) In step-i, data qubits on the logical qubit 1 are initialised in the state $\ket{+}$, but the rightmost column is skipped. In step-ii and step-iii, stabilisers are measured for $h\sim d$ rounds. For the first $h-1$ rounds (step-ii), the rightmost column of the logical qubit 1 is skipped. For the $h$-th round (step-iii), the state of the leftmost column of the logical qubit 2 is transferred to the rightmost column of the logical qubit 1. In step-iv, data qubits of the logical qubit 2 are measured in the $+$/$-$ basis. (b) The circuit of X-stabiliser measurements on columns B, D and E. Qubits without labels are shuttle qubits, and qubits with labels are data qubits in corresponding columns. The part of the circuit in the orange shadow creates a Bell state on the top shuttle qubit (shared by columns B and C) and the bottom shuttle qubit (shared by columns D and E), which replaces the CNOT gate between two shuttle qubits in Fig.~\ref{fig:circuit}(c). (c) The circuit of Z-stabiliser measurements on columns A, B and D. The part of the circuit in the orange shadow creates a Bell state on the top shuttle qubit (shared by columns A and B) and the bottom shuttle qubit (shared by columns C and D), which replaces the CNOT gate between two shuttle qubits in Fig.~\ref{fig:circuit}(d). (d) The circuit of Z-stabiliser measurements on columns A, B and D, and the state of the D-column data qubit is simultaneously transferred to the C-column data qubit. The part of the circuit in the orange shadow is a quantum teleportation that transfers the state of the D-column data qubit to the C-column data qubit. Using circuits in (b,c), stabilisers across the boundary of two logical qubits can be measured, and other stabilisers are measured according to circuits in Fig.~\ref{fig:circuit}(c,d). For the $h$-th round stabiliser measurements, X stabilisers are measured firstly, then Z stabilisers are measured using the circuit in (d).
}
\label{fig:state_transfer}
\end{figure*}

Details of our protocols of the surface-code logical CNOT gate, Hadamard gate and state transfer are shown in Fig.~\ref{fig:CNOT}, Fig.~\ref{fig:Hadamard} and Fig.~\ref{fig:state_transfer}, respectively.

\end{document}